\documentclass{cernrep} 
\usepackage{texnames}
\usepackage[T1]{fontenc}
\usepackage[bookmarks, colorlinks=true, linktoc=page, pdftex, linkcolor=blue, citecolor=blue, urlcolor=blue]{hyperref}
\def\unsty{\ \mathrm}

\usepackage{fancyhdr}
\fancyhfoffset{4 mm}
\fancypagestyle{ARTTITLE}{%
\fancyhf{} % clear all header and footer fields
\lhead{\small{Proceedings of the CERN--Accelerator--School course: \it{Introduction to Accelerator Physics}, courses 2019 and beyond}}
\lfoot{Available online at \url{https://cas.web.cern.ch/previous-schools}}
\rfoot{\thepage\hspace*{3mm}}

}

\begin{document}
\title{Introduction to Non-linear Longitudinal Beam Dynamics}
 
\author {H. Damerau}

\institute{CERN, Geneva, Switzerland}

\begin{abstract}

The interaction of a charged particle beam with radio-frequency (RF) systems in most linear or circular accelerators is an non-linear process. The large longitudinal electric fields for acceleration and longitudinal beam manipulations can only be generated thanks to the resonant build-up of the field in a high quality oscillator, the RF cavity, driven by a sinusoidal and hence inherently non-linear excitation. The course gives an introduction to linear and non-linear longitudinal beam dynamics, deriving the equations of motion, as well as the RF potential and the Hamiltonian of the longitudinal beam dynamics. Profiting from the non-linear dynamics, longitudinal beam manipulations to control bunch length, distance with multiple RF systems are shown as examples. Additionally, the distribution of the synchrotron frequencies of the particles in a bunch can be modelled thanks to the non-linearity of the applied RF voltage. Double- or even multi-harmonic RF systems are a powerful technique to improve the longitudinal stability of high-intensity beams.

\end{abstract}


\maketitle % this produces the title block
\thispagestyle{ARTTITLE}
 
\section{Introduction}

Large longitudinal electrical fields required for the acceleration of charged particles are achieved by resonant systems. RF cavities as oscillators with high quality factors assure this excitation of such fields. As a consequence of the small bandwidth, most RF systems in particle accelerators are operated using sinusoidal RF voltages, which are inherently non-linear.

The approach to longitudinal beam dynamics assuming a linear RF voltage is nonetheless a very relevant approximation in many cases. The contribution of the non-linearity of the RF voltage to longitudinal beam dynamics is nonetheless essential to keep high-intensity beams stable in many circular accelerators.

Additional non-linearity in the form of one or more extra RF systems, again with sinusoidal voltages, is actually often introduced intentionally. This allows to modify the longitudinal beam parameters like bunch length, bunch spacing, as well as the longitudinal stability most efficiently.

A brief, general overview to the longitudinal beam dynamics in synchrotrons describing the interaction between single particles or bunches with the RF systems is introduced for the case an arbitrary normalized amplitude. Attention is then given to the model case of a linear RF waveform. This special case not only allows to gain insight into the longitudinal motion in the centre of basically any bucket, but also outlines the differences with respect to the non-linear case.

Examples for longitudinal manipulations to adjust the bunch parameters show the versatility of double- and multi-harmonic RF systems. The synchrotron frequency distribution of the individual particles in a bunch has a strong impact on its longitudinal stability. Techniques are introduced to improve the stability by making use of non-linear voltage, notably a combination of two sinusoidal RF voltages, the double-harmonic RF system.

\section{Longitudinal beam dynamics}
\label{sec:LongitudinalBeamDynamics}

The momentum, $p$ of a relativistic particle in an accelerator can be written in the form of
\begin{equation}
    p = m v = m_0 \beta \gamma c \, ,
\end{equation}
where $m_0$ is the rest mass of the particle, $\beta = v/c$ its relative velocity ratio with respect to the speed of light in vacuum, $c$ and $\gamma = m/m_0$ the relativistic mass factor.

Synchrotrons profit from the fact that particles pass through the accelerating systems for some thousands to millions of turns. The momentum change per turn is therefore moderate with respect to the total particle momentum. When the particle passes through the RF station once, it can be safely assumed that the momentum gain or loss remains small compared to the total one. The momentum change, $dp/dt$ is hence expressed in its discrete form. Within a single turn the momentum changes by $\Delta p$, and $\Delta t$ becomes the revolution period, $T_\mathrm{rev} = 1/f_\mathrm{rev} = \beta c/(2 \pi R)$. The momentum change with time can be transformed according to 
\begin{equation}
    \dot{p} = \frac{dp}{dt} \simeq \frac{\Delta p}{\Delta t} = \frac{\Delta p}{T_\mathrm{rev}} = \frac{m_0 \beta c^2}{2 \pi R} \Delta (\beta \gamma) = \frac{m_0 \beta c^2}{2 \pi R} (\beta \Delta \gamma + \gamma \Delta \beta) = \frac{\Delta E}{2 \pi R} \, .
    \label{eqn:MomentumDerivativeToEnergyGainPerTurn}
\end{equation}
To replace $\Delta \beta$ in Eq.~(\ref{eqn:MomentumDerivativeToEnergyGainPerTurn}) by $\Delta \gamma$ the relativistic relations
\begin{equation*}
    \beta (\gamma)= \sqrt{1-\frac{1}{\gamma^2}}  \hspace{1em} \text{and its derivative} \hspace{1em} \frac{\Delta \beta}{\Delta \gamma} \simeq \frac{d \beta}{d \gamma} = \frac{1}{\beta \gamma^3}
\end{equation*}
have been applied.

For the synchronous particle as a reference, momentum and energy changes can be related to the derivative of the bending field during acceleration. A particle with the mass $m$ and the charge $q$ moving with the velocity $v$ in a perpendicular magnetic field of the strength $B$ experiences the Lorentz force, $F_\mathrm{L} = q v B$. This force must compensate the centripetal force, $F_\mathrm{Z} = m v^2/\rho$, which translates to
\begin{equation}
    F_L = F_Z \hspace{1em} \Longrightarrow \hspace{1em} p = q \rho B \, ,
\end{equation}
where $\rho$ represents the bending radius in the dipole magnets. Using Eq.~(\ref{eqn:MomentumDerivativeToEnergyGainPerTurn}) the energy gain or loss can then be written as
\begin{equation}
    \Delta E_\mathrm{turn} = 2 \pi q \rho R \dot{B} \, .
    \label{eqn:energyChangePerTurnDueToBdot}
\end{equation}

Any energy change, $\Delta E_\mathrm{turn}$ must be compensated by the RF system to keep the particle on a central orbit which defines the voltage, $V g(\phi)$ during its passage though the cavities and hence
\begin{equation}
    \Delta E_\mathrm{turn} = q V g (\phi) \, ,
    \label{eqn:energyChangePerTurnDueToVoltage}
\end{equation}
where $V$ is the absolute peak voltage amplitude and $g(\phi)$ its normalized amplitude.

Combining Eq.~(\ref{eqn:energyChangePerTurnDueToBdot}) with Eq.~(\ref{eqn:energyChangePerTurnDueToVoltage}) and assuming a sinusoidal RF amplitude, $g(\phi) = \sin(\phi)$ results in the conventional relation defining the synchronous phase
\begin{equation}
    \sin \phi_\mathrm{S} = \frac{qV}{2 \pi q \rho R \dot{B}}
\end{equation}
at which the so-called synchronous particle receives exactly the energy change required to stay on the central orbit with the increasing or decreasing bending field.

For any other particle with a phase offset of $\Delta \phi = \phi - 
\phi_\mathrm{S}$ with respect to the synchronous particle the time derivative of its energy offset, $\Delta E$ is expressed as
\begin{equation}
    \frac{d}{dt} \Delta E = q V \frac{\omega_\mathrm{rev}}{2 \pi} \left[ g(\phi) - g( \phi_\mathrm{S} ) \right] \, ,
    \label{eqn:hamiltonEquationDeltaE}
\end{equation}
where $\omega_\mathrm{rev} = 2\pi f_\mathrm{rev}$ is the angular revolution frequency.

The relationship which links frequency and momentum offset in a synchrotron is called phase slip factor and is defined as\footnote{The phase slip factor is sometimes defined with the opposite sign, e.g.~\cite{bib:dome1984}.}
\begin{equation}
\eta = - \frac{\Delta \omega/\omega}{\Delta p/p} = \alpha_c - \frac{1}{\gamma^2} = \frac{1}{\gamma_\mathrm{tr}^2} - \frac{1}{\gamma^2} \, ,
\end{equation}
with the momentum compaction factor $\alpha_c = (\Delta R/R)/(\Delta p/p) = 1/\gamma_\mathrm{tr}^2$. The momentum compaction factor is a property of the magnet lattice~\cite{bib:bovet1970}. The relative frequency deviation $\Delta \omega/\omega$ can be either with respect to the revolution, $\Delta \omega_\mathrm{rev}/\omega_\mathrm{rev}$ or the RF frequency $\Delta \omega_\mathrm{RF}/\omega_\mathrm{RF}$, where $\omega_\mathrm{RF} = h \omega_\mathrm{rev}$. The ratio between RF and revolution frequency is called harmonic number, $h = \omega / \omega_\mathrm{rev}$.

The phase change of a particle due to a frequency offset is given by~\cite{bib:pirkl1995, bib:montague1977}
\begin{equation}
    \frac{d}{dt} \phi = - h \Delta \omega_\mathrm{rev} \, ,
    \label{eqn:frequencyOffsetToPhaseDerivative}
\end{equation}
where the minus sign is a consequence of the choice of the RF phase, $\phi$ as a variable. A particle with positive revolution frequency offset arrives too early in the cavity and experiences the RF voltage at a lower phase with respect to the synchronous reference particle.

Replacing $\Delta p/p = 1/\beta^2 \Delta E/E$ and using Eq.~(\ref{eqn:frequencyOffsetToPhaseDerivative}) results in~\cite{bib:dome1984} 
\begin{equation}
\frac{d}{dt} \phi =  h \omega_\mathrm{rev} \frac{\eta}{E \beta^2} \Delta E = \frac{h \eta \omega_\mathrm{rev}}{pR} \left( \frac{\Delta E}{\omega_\mathrm{rev}} \right) \, .
\label{eqn:hamiltonEquationDeltaPhi}
\end{equation}
Equations~(\ref{eqn:hamiltonEquationDeltaE}) and (\ref{eqn:hamiltonEquationDeltaPhi}) form a set of Hamilton's equations as their structure matches
\begin{equation}
    \frac{dq}{dt} = \frac{ \partial H}{\partial p} \hspace{1em} \text{and} \hspace{1em} \frac{dp}{dt} = - \frac{ \partial H}{\partial q} \, ,
\end{equation}
with $\phi$ and $\Delta E/ \omega_\mathrm{rev}$ as canonically conjugated variables $q$ and $p$:
\begin{eqnarray}
\frac{d}{dt} \phi & = & \frac{h \eta \omega_\mathrm{rev}}{pR} \left( \frac{\Delta E}{\omega_\mathrm{rev}} \right)  \label{eqn:firstHamiltonEquation} \\
\frac{d}{dt} \left( \frac{\Delta E}{\omega_\mathrm{rev}} \right) & = & \frac{q V}{2 \pi} \left[ g(\phi) - g( \phi_\mathrm{S} ) \right] \label{eqn:secondHamiltonEquation} \, .
\end{eqnarray}
The choice of variables $\phi$ and $\Delta E / \omega_\mathrm{rev}$ defines the explicit form of the Hamiltonian, and it is convenient to calculate invariant areas in the longitudinal phase space using these variables.
Combining both equations Eq.~(\ref{eqn:firstHamiltonEquation}) and Eq.~(\ref{eqn:secondHamiltonEquation}) leads to the Hamiltonian of the synchrotron motion
\begin{equation}
    H \left( \phi , \frac{\Delta E}{\omega_\mathrm{rev}} \right) = \frac{1}{2} \frac{h \eta \omega_\mathrm{rev}}{p R} \left( \frac{\Delta E}{\omega_\mathrm{rev}} \right)^2 - \frac{qV}{2 \pi} \left[ \int g(\phi) d \phi - g(\phi_\mathrm{S}) \phi \right] \, .
\end{equation}
It describes the behaviour of a particle in the coordinates of phase, $\phi$, and energy offset $\Delta E$. For any given particle trajectory in the longitudinal $\phi$-$\Delta E/\omega_\mathrm{rev}$-phase space the value of the Hamiltonian varies very slowly with respect to the dynamics of the system.

The second term of the Hamiltonian is identified as the normalized potential
\begin{equation}
    W (\phi) = \frac{1}{\cos \phi_\mathrm{S}} \left[ \int g(\phi) d \phi - g(\phi_\mathrm{S}) \phi \right] \, ,
    \label{eqn:normalizedPotential}
\end{equation}
which is basically the integral of the RF voltage. The Hamiltonian can hence be written as
\begin{equation}
    H \left( \phi , \frac{\Delta E}{\omega_\mathrm{rev}} \right) = \frac{1}{2} \frac{h \eta \omega_\mathrm{rev}}{p R} \left( \frac{\Delta E}{\omega_\mathrm{rev}} \right)^2 - \frac{qV\cos\phi_\mathrm{S}}{2 \pi} \, W (\phi) \, .
    \label{eqn:GeneralHamiltonian}
\end{equation}

The general equation of motion is obtained by differentiating Eq.~(\ref{eqn:firstHamiltonEquation}) in time and substituting the result in Eq.~(\ref{eqn:secondHamiltonEquation}). This general equation can be written as
\begin{equation}
    \frac{d^2}{dt^2} \phi + \frac{ \omega_\mathrm{S}^2 }{ \cos \phi_\mathrm{S} } \left[ g(\phi) - g(\phi_\mathrm{S}) \right] = 0 \hspace{1em} \text{with} \hspace{1em} \omega_\mathrm{S}^2 = - \frac{ h \eta \omega_\mathrm{rev} q V \cos \phi_\mathrm{S} }{ 2 \pi p R} \, ,
    \label{eqn:GeneralEquationOfMotionAndSynchrotronFrequency}
\end{equation}
where $\omega_\mathrm{S}$ is the angular frequency of the longitudinal oscillation in phase and energy of the particle.

With the help of Eq.~(\ref{eqn:firstHamiltonEquation}) the energy offset variable can be changed from $\Delta E/\omega_\mathrm{rev}$ to $\dot{\phi}/\omega_\mathrm{S}$, and the Hamiltonian Eq.~(\ref{eqn:GeneralHamiltonian}) for an arbitrary RF voltage function $g(\phi)$ can be rewritten in the elegant form of
\begin{equation}
    H \left( \phi, \frac{\dot{\phi}}{\omega_\mathrm{S}} \right) = \frac{1}{2} \left( \frac{ \dot{\phi} }{ \omega_\mathrm{S}} \right)^2 + W(\phi)
    \label{eqn:GeneralHamiltonianPhiPhiDot}
\end{equation}
again with the normalized potential according to Eq.~(\ref{eqn:normalizedPotential}). Compared to Eq.~(\ref{eqn:GeneralHamiltonian}) the Hamiltonian has been divided by the factor of $k = - q V \cos \phi_\mathrm{S} / (2 \pi )$. The stable phase angle $\phi_\mathrm{S}$ is not necessarily unambiguously defined as an angle but indicates the energy gain or loss per turn of the synchronous particle.

\section{Linear longitudinal dynamics}

\begin{figure}[htb]
	\centering\includegraphics[width=.4\linewidth]{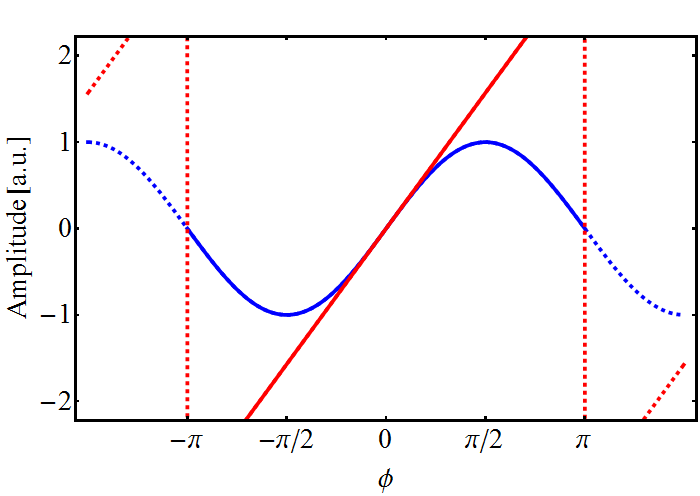}
	\caption{Sinusoidal (blue) and linear (red) RF amplitude.}
	\label{fig:SinusoidalLinearVoltage}
\end{figure}

The longitudinal dynamics of a particle under the influence of a linear RF field is mainly a simplified mathematical model to describe the motion in the linear regions of a non-linear RF amplitude. Figure~\ref{fig:SinusoidalLinearVoltage} compares sinusoidal, $g(\phi) = \sin \phi$ and linear, $g(\phi) = \phi$ normalised RF amplitudes.
For the case of no net acceleration, $\phi_\mathrm{S} = 0$ and $g(\phi_\mathrm{S})=0$, the synchronous particle is located at the zero crossing of the RF voltage, where linear and sinusoidal RF amplitudes are almost identical. It is worth noting that the saw-tooth waveform has a large frequency spectrum which technically makes the generation of large voltages unfavourable, and it is therefore of only limited practical interest for implementation. 

For a linear RF amplitude the Hamilton equations Eq.~(\ref{eqn:firstHamiltonEquation}) and Eq.~(\ref{eqn:secondHamiltonEquation}) simplify to 
\begin{eqnarray}
\frac{d}{dt} \phi & = & \frac{h \eta \omega_\mathrm{rev}}{pR} \left( \frac{\Delta E}{\omega_\mathrm{rev}} \right)  \label{eqn:firstHamiltonEquationLinearVoltage} \\
\frac{d}{dt} \left( \frac{\Delta E}{\omega_\mathrm{rev}} \right) & = & \frac{q V}{2 \pi} \phi \label{eqn:secondHamiltonEquationLinearVoltage} \, .
\end{eqnarray}
The corresponding Hamiltonian for this linear longitudinal particle motion can be written as
\begin{equation}
     H \left( \phi , \frac{\Delta E}{\omega_\mathrm{rev}} \right) = \frac{1}{2} \frac{h \eta \omega_\mathrm{rev}}{p R} \left( \frac{\Delta E}{\omega_\mathrm{rev}} \right)^2 - \frac{1}{2} \frac{qV}{2 \pi} \phi^2 \, .
     \label{eqn:HamiltonianLinearVoltage}
\end{equation}

The explicit form of this Hamiltonian can again be significantly simplified by changing the variable describing the energy offset of the particle. Equation~(\ref{eqn:secondHamiltonEquationLinearVoltage}) actually translates this energy offset, $\Delta E/\omega_\mathrm{rev}$ into a drift velocity, $d\phi/dt = \dot{\phi}$ with respect to the synchronous particle. In the new set of variables $\phi$ and $\dot{\phi}/\omega_\mathrm{S}$ the explicit form of the Hamiltonian becomes
\begin{equation}
    H \left( \phi , \frac{\dot{\phi}}{\omega_\mathrm{S}} \right) = \frac{1}{2} \left( \frac{\dot{\phi}}{\omega_\mathrm{S}} \right)^2 + \frac{1}{2} \phi^2 \, ,
    \label{eqn:HamiltonianLinearVoltageNormalized}
\end{equation}
applying the same normalization factor, $k$ as for Eq.~(\ref{eqn:GeneralHamiltonianPhiPhiDot}).

\begin{figure}[htb]
	\centering
	\includegraphics[width=0.329\linewidth]{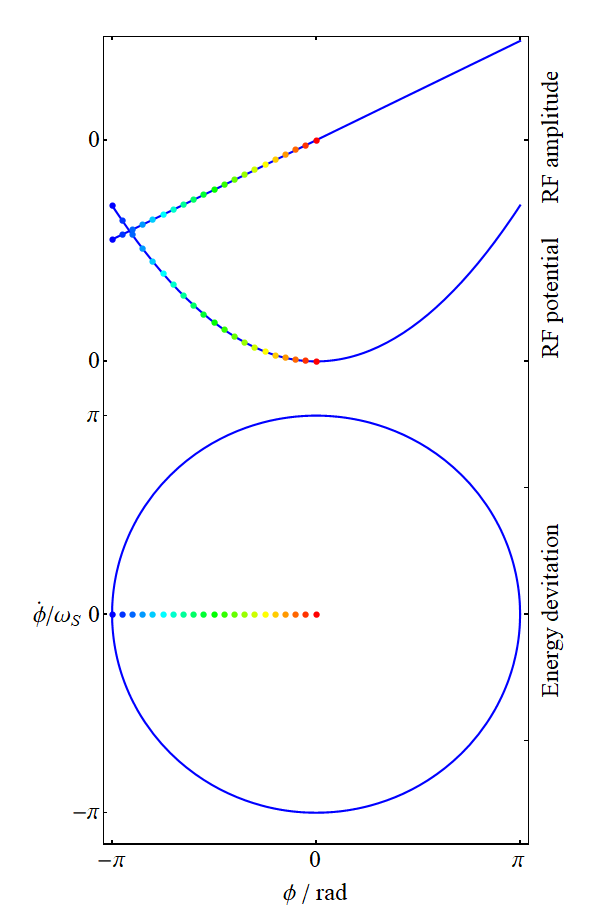}
	\includegraphics[width=0.329\linewidth]{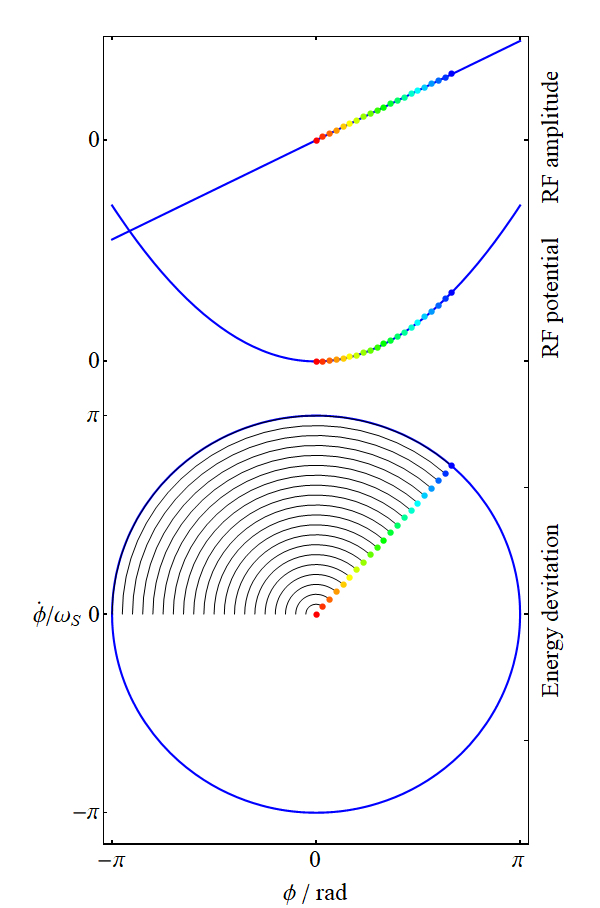}
	\includegraphics[width=0.329\linewidth]{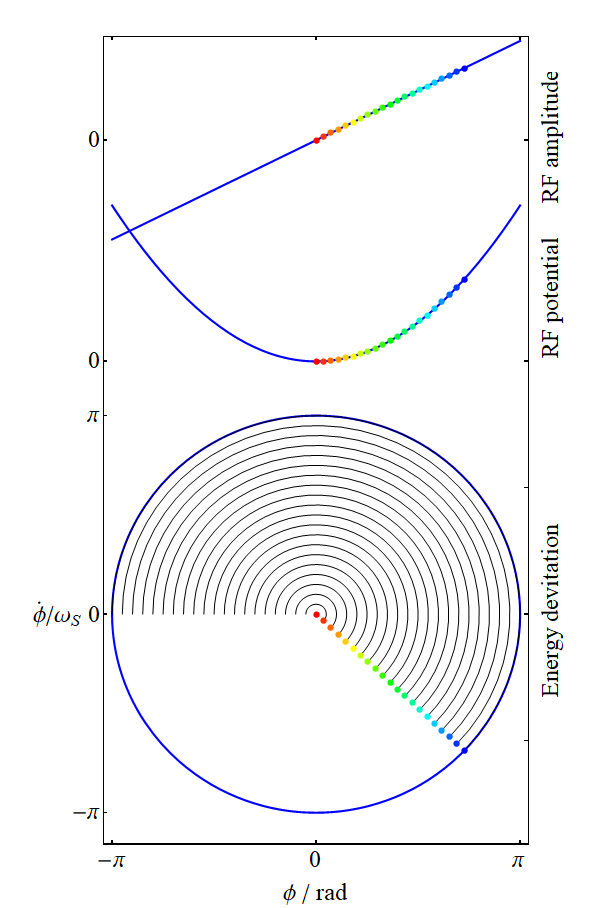}
	\caption{Amplitude (top), resulting potential (middle) and normalized longitudinal phase space (bottom) in coordinates of $\phi$ and $\dot{\phi}/\omega_\mathrm{S}$. The phase space trajectories are just circles and the oscillation frequency is the same for all particles.}
	\label{fig:VoltagePotentialPhaseSpaceLinearVoltage}
\end{figure}

The relation describes the equation of a circle, $r^2 = x^2 + y^2$. Therefore, the trajectories of all particles under the influence of a linear RF amplitude can be represented as circles in the longitudinal $\phi$-$\dot{\phi}/\omega_\mathrm{S}$-phase space~(Fig.~\ref{fig:VoltagePotentialPhaseSpaceLinearVoltage}):
\begin{equation*}
    \dot{\phi} (\phi) = \omega_\mathrm{S} \sqrt{2 H_\mathrm{trajectory} -  \phi^2} \, .
\end{equation*}
The RF potential is then a parabolic function, $W \propto \phi^2$.
As indicated by the trajectories (bottom) all particles moreover have the same synchrotron frequency, independent of their energy or phase deviation with respect to the bucket center.

\section{Sinusoidal RF voltage}

For the most common case of a single sinusoidal RF voltage the term $V\phi$ in Eq.~(\ref{eqn:secondHamiltonEquationLinearVoltage}) is replaced by $V \sin \phi$ which introduces a simple non-linearity. As explained above, sinusoidal RF voltages are most common in accelerators with conventional RF systems since large voltages are easily achieved in the resonant RF cavities. The Hamilton equations for this case become
\begin{eqnarray}
\frac{d}{dt} \phi & = & \frac{h \eta \omega_\mathrm{rev}}{pR} \left( \frac{\Delta E}{\omega_\mathrm{rev}} \right) \\
\frac{d}{dt} \left( \frac{\Delta E}{\omega_\mathrm{rev}} \right) & = & \frac{q V}{2 \pi} \left( \sin \phi - \sin \phi_\mathrm{S} \right) \label{eqn:secondHamiltonEquationSinusoidalVoltage}
\end{eqnarray}
and can be combined to the Hamiltonian
\begin{equation}
    H \left( \phi , \frac{\Delta E}{\omega_\mathrm{rev}} \right) = \frac{1}{2} \frac{h \eta \omega_\mathrm{rev}}{p R} \left( \frac{\Delta E}{\omega_\mathrm{rev}} \right)^2 + \frac{qV}{2 \pi} \left[ \cos\phi - \cos \phi_\mathrm{S} + (\phi - \phi_\mathrm{S}) \sin \phi_\mathrm{S} \right] \, .
\end{equation}
This standard expression, and its detailed analysis of stable parameter regions in the phase space of $\Delta \phi$ and $\Delta E/\omega_\mathrm{rev}$, can be found in a large number of text books and review articles~\cite{bib:montague1977,bib:pirkl1995,bib:tecker2006,bib:tecker2019}.

To study the behaviour of particles close to the centre of the bunch, it is convenient to introduce $\Delta \phi = \phi - \phi_\mathrm{S}$ which brings the Hamiltonian in the form
\begin{equation}
    H \left( \phi , \frac{\Delta E}{\omega_\mathrm{rev}} \right) = \frac{1}{2} \frac{h \eta \omega_\mathrm{rev}}{p R} \left( \frac{\Delta E}{\omega_\mathrm{rev}} \right)^2 + \frac{qV}{2 \pi} \left[ \cos(\phi_\mathrm{S} + \Delta \phi) - \cos \phi_\mathrm{S} + \Delta \phi \sin \phi_\mathrm{S} \right] \, . \label{eqn:hamiltonianSinusoidalVoltage}
\end{equation}
In the limit of $\Delta \phi \ll 1$, valid for particles close to the synchronous one, the first part of the potential term can be rearranged using
\begin{eqnarray}
    \cos(\phi_\mathrm{S} + \Delta \phi) & = & \cos(\phi_\mathrm{S}) \cos \Delta \phi - \sin \phi_\mathrm{S} \sin \Delta \phi \\
    & \simeq & \cos \phi_\mathrm{S} \left( 1- \frac{1}{2} \Delta \phi^2 \right) - \sin \phi_\mathrm{S} \Delta \phi
\end{eqnarray}
such that Eq.~(\ref{eqn:hamiltonianSinusoidalVoltage}) is approximated to
\begin{equation}
    H \left( \Delta \phi , \frac{\Delta E}{\omega_\mathrm{rev}} \right) \simeq \frac{1}{2} \frac{h \eta \omega_\mathrm{rev}}{p R} \left( \frac{\Delta E}{\omega_\mathrm{rev}} \right)^2 - \frac{1}{2} \frac{qV}{2 \pi} \cos \phi_\mathrm{S} \Delta \, \phi^2 \, . \label{eqn:hamiltonianSinusoidalVoltageSmallAmplitudes}    
\end{equation}
As in Eq.~(\ref{eqn:HamiltonianLinearVoltage}) the explicit dependency of $\Delta \phi$ and $\Delta E / \omega_\mathrm{rev}$ is again quadratic, indicating that trajectories with constant $H ( \Delta \phi , \Delta E / \omega_\mathrm{rev} )$, as well as small phase and energy deviations with respect to the synchronous particle at the stable phase are elliptic (or circular, depending on the normalization of coordinates). Each particle performs synchrotron oscillations in the longitudinal phase space, constantly exchanging phase and energy offset, with the synchrotron frequency according to Eq.~(\ref{eqn:GeneralEquationOfMotionAndSynchrotronFrequency}).

\begin{figure}[htb]
	\centering\includegraphics[width=.4\linewidth]{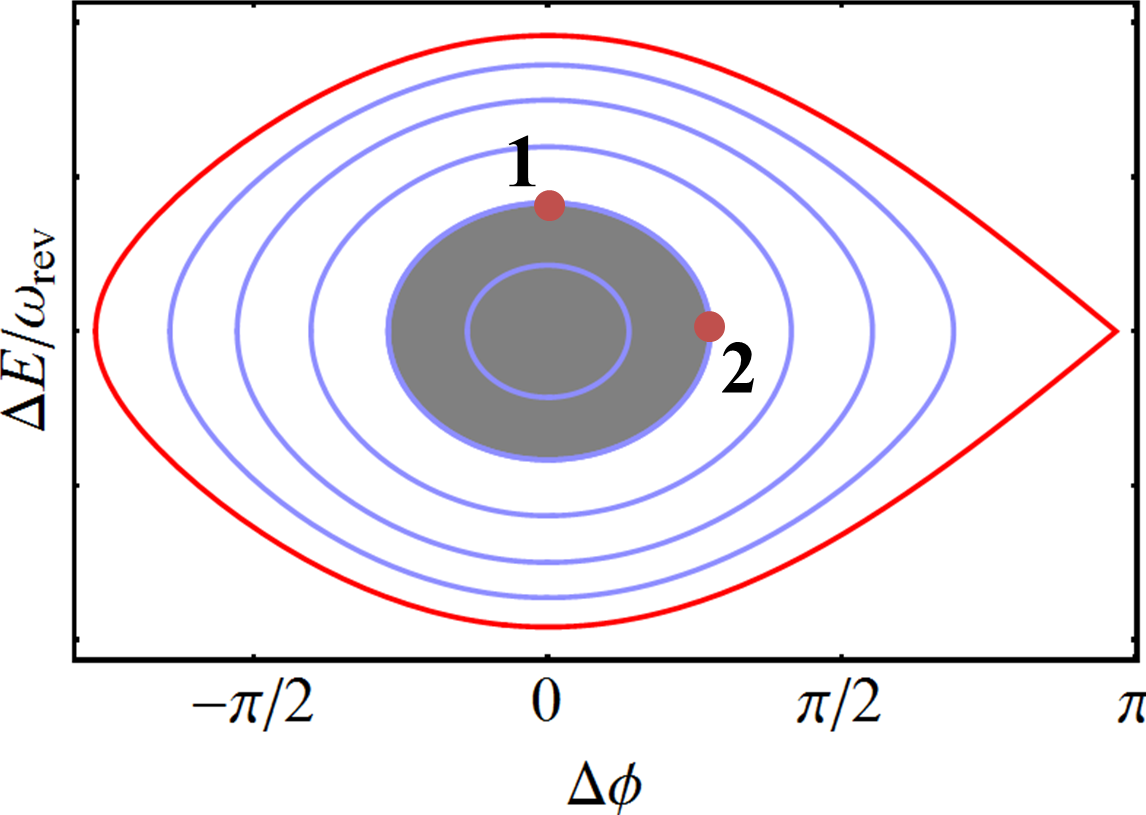}
	\caption{The gray shaded area occupied by particles indicates the longitudinal emittance. The separatrix (red) indicates the limit of closed trajectories.}
	\label{fig:AcceleratingDecelratingBucketCropped}
\end{figure}

Figure~\ref{fig:AcceleratingDecelratingBucketCropped} illustrates the longitudinal phase space for the case of an accelerating bucket.
For any given trajectory the value of the Hamiltonian is constant. The extremes (Fig.~\ref{fig:AcceleratingDecelratingBucketCropped}) of two particles on the same trajectory are the case zero phase (1) or energy deviation (2). For these particles the Hamiltonian is expressed as
\begin{eqnarray}
    H \left( \Delta \phi = 0, \frac{\Delta E}{\omega_\mathrm{rev}} \right) = H_\mathrm{trajectory}  & = & \phantom{-} \frac{1}{2} \frac{h \eta \omega_\mathrm{rev}}{p R} \left( \frac{\Delta E}{\omega_\mathrm{rev}} \right)^2 \hspace{1em} \mathrm{and} \\
    H \left( \Delta \phi, \frac{\Delta E}{\omega_\mathrm{rev}} = 0 \right) = H_\mathrm{trajectory} & = & - \frac{1}{2} \frac{qV}{2 \pi} \cos \phi_\mathrm{S} \Delta \, \phi^2 \, .
\end{eqnarray}

The longitudinal emittance, $\varepsilon_\mathrm{l}$ is the surface in the longitudinal phase space occupied by the ensemble of all particles particles on stable trajectories around the synchronous particle, the so called bunch. In the example shown in Fig.~\ref{fig:AcceleratingDecelratingBucketCropped} it is represented by the gray shaded area. Choosing a canonically conjugated set of variables its surface is preserved, just like an incompressible fluid, during acceleration or RF manipulations. For the common choice of time and energy offset, the unit of the longitudinal emittance becomes $[\varepsilon_\mathrm{l}] = [\pi \Delta \tau \Delta E] = \mathrm{eVs}$. The duration from centre of the bunch to the maximum phase excursion of the encircling trajectory is $\Delta \tau$ (half the total bunch length), and the height of the trajectory is $\Delta E$ (half the total energy width).

For any given trajectory, independent of the non-linearity of the RF waveform, the longitudinal emittance can be calculated according to
\begin{equation}
    \varepsilon_\mathrm{l} = \frac{2}{h \omega_\mathrm{rev}} \int_{\Delta \phi_\mathrm{i}}^{\Delta \phi_\mathrm{f}} \Delta E (\Delta \phi) \, d(\Delta \phi) \, ,
\end{equation}
which is the surface within the surrounding trajectory $\pm \Delta E (\Delta \phi)$ of the bunch.

\section{Controlling bunch parameters with multi-harmonic RF systems}

Due the the resonant nature of conventional RF systems, this section concentrates on the combination of an accelerating voltage delivered by multiple RF stations, each of which generates a sinusoidal voltage. In the most simple extension of two harmonic RF systems, the voltage $V \sin \phi$ is replaced by $V_1 \sin \phi + V_2 \sin( h_2/h_1 \phi + \phi_1) ]$, where $h_2/h_1$ is the harmonic number ratio and $V_2$ is the voltage of the second harmonic RF system. Already for an RF voltage composed of two sinusoidal amplitudes, analytical estimates of the bucket parameters are only possible for few special cases. Generally, the trajectories can be calculated numerically using Eq.~(\ref{eqn:GeneralHamiltonianPhiPhiDot}).

\subsection{Bunch length manipulations}

\begin{figure}[htb]
	\centering
	\includegraphics[width=0.45\linewidth]{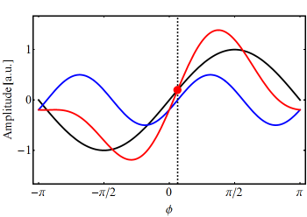}
	\includegraphics[width=0.45\linewidth]{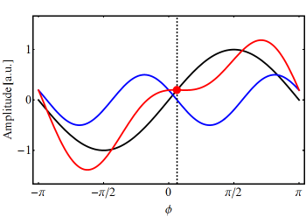}
	\caption{Sum of two RF voltages (red) with a harmonic ratio of $h_2/h_1=2$ (black and blue). The amplitude of the higher harmonic system is half the amplitude of the fundamental one. The effect on the beam depends strongly on whether the voltages are at the same phase (left) or in opposite phase (right) at the centre of the bunch (dashed line).}
	\label{fig:doubleHarmonicBunchShorteningLengtheningModeExample}
\end{figure}

Two examples of a double harmonic RF voltage, with $h_2/h_1=2$ and $V_2/V_1=0.5$ are illustrated in Fig.~\ref{fig:doubleHarmonicBunchShorteningLengtheningModeExample}.
When both the sinusoidal amplitudes are in phase at the position of the bunch, the local voltage gradient is increased, decreasing the bunch length. This configuration is called bunch shortening mode and results in a larger peak current. Nonetheless this case is has a strong impact on longitudinal stability as shown in Sec.~\ref{sec:bunchShorteningMode}. In the opposite case, when both RF systems are in counter-phase in the bunch lengthening mode, the local voltage gradient is decreased, which stretches the bunch and reduces the peak longitudinal line density.

\begin{figure}[htb]
	\centering
	\includegraphics[width=0.42\linewidth]{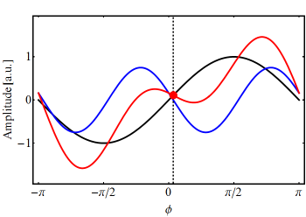}
	\includegraphics[width=0.48\linewidth]{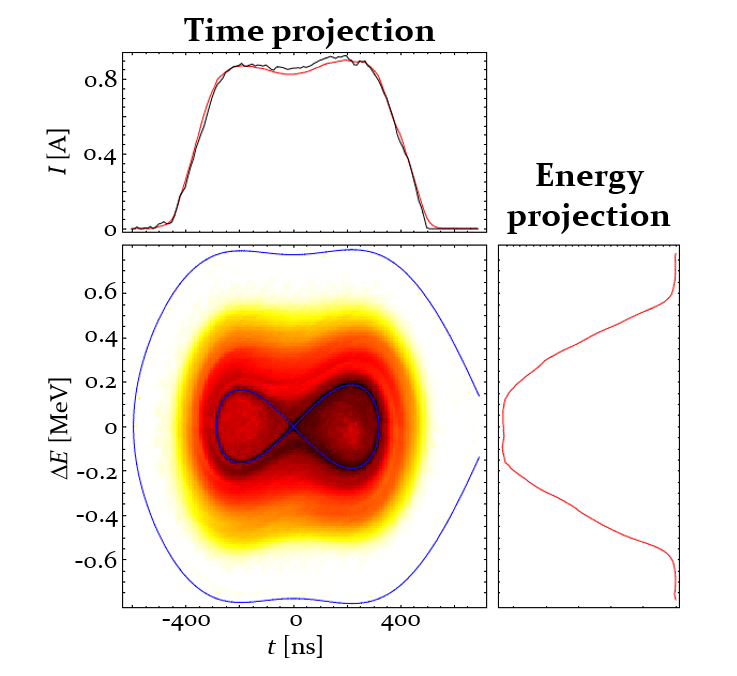}
	\caption{The RF voltage during the low energy part of the acceleration cycle in the CERN Proton Synchrotron Booster (PSB) used two harmonics a $h_1 = 1$ and $h_2 = 2$. The measured bunch distribution under these conditions (right) is stretched such that the time projection, the bunch profile has two peaks around the depleted bunch center.}
	\label{fig:doubleHarmonicBunchLengtheningModeExamplePSB}
\end{figure}

The most important application of the bunch lengthening mode is the reduction of the instantaneous current is the reduction of the space charge effect~\cite{bib:ferrario2014}. The lower the instantaneous beam current, the fewer particles are squeezed into a given part of the circumference, hence reducing the electric forces between particles. Figure~\ref{fig:doubleHarmonicBunchLengtheningModeExamplePSB} shows the RF voltage used in the Proton Synchrotron Booster (PSB) at CERN during the low energy part of the acceleration ramp (left) and the resulting longitudinal distribution of the bunch.
The important bunch flattening and depletion of the distribution in the centre due to the inverted voltage gradient reduces the maximum longitudinal line density and significantly reduces the space charge effect.

\subsection{Adjustment of the bunch distance}

Not only the bunch length, but also the distance of the bunches, normally an integer multiple of the RF wavelength, can be controlled with double-harmonic RF systems. The circumference of the Proton Synchrotron (PS) at CERN is exactly four times larger than the four sandwiched rings of its injector, the aforementioned PSB. An RF harmonic of $h_\mathrm{PS} = 8$ in the the PS thus corresponds to two bunches with a spacing of half a turn in the corresponding single-harmonic RF voltage at $h_\mathrm{PSB} = 2$.

Adding some small RF voltage at the revolution frequency, $h_\mathrm{PSB} = 1$, allows to control the bunch spacing and adapt it to a different harmonic number in the PS. With this technique the circumference ratio is virtually moved to $h_\mathrm{PS}/h_\mathrm{PSB} = 7/2$ for the transfer of beams for the Large Hadron Collider (LHC) between PSB and PS~\cite{bib:hancock2010}.

\begin{figure}[htb]
	\centering\includegraphics[width=.4\linewidth]{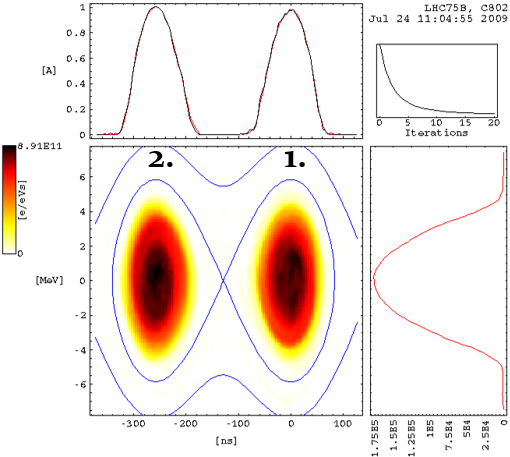}
	\caption{Reconstructed longitudinal distribution of two bunches in the main RF harmonic of $h_\mathrm{PSB} = 2$ under the influence of additional voltage at $h_\mathrm{PSB} = 1$. Instead of a spacing of $286\unsty{ns}$ for the single-harmonic case, the distance between the bunches is reduced to about $245\unsty{ns}$.}
	\label{fig:singleBatchTransferLongitudinalPhaseSpace}
\end{figure}

Figure~\ref{fig:singleBatchTransferLongitudinalPhaseSpace} shows the measured longitudinal phase space, as well as time and energy projections with $h_\mathrm{PSB} = 2$ complemented by additional voltage at $h_\mathrm{PSB} = 1$.
The revolution frequency of the beam in the PSB at a kinetic energy of $E_\mathrm{kin} = 1.4\unsty{GeV}$ is $1.75\unsty{MHz}$ and the natural bunch spacing would be $\tau_{h=2} = T_\mathrm{rev,PSB}/2 =  286\unsty{ns}$. Applying RF voltage at the revolution frequency reduces the short bunch spacing in the PSB to $\tau_\mathrm{short} = 245\unsty{ns}$. Effectively the bunches are slightly displaced in azimuth with respect to their diametrically opposite position, and the long distance between to bunches becomes $\tau_\mathrm{short} = 327\unsty{ns}$, exactly the period of the RF voltage for the capture into $h_\mathrm{PS} = 7$-buckets in the PS. Since the manipulation does not change the revolution frequencies, the bunch spacings must follow the relation to $T_\mathrm{rev,PSB} = 2 \tau_\mathrm{h=2} = \tau_\mathrm{short} + \tau_\mathrm{long}$.

\begin{figure}[htb]
	\centering\includegraphics[width=.6\linewidth]{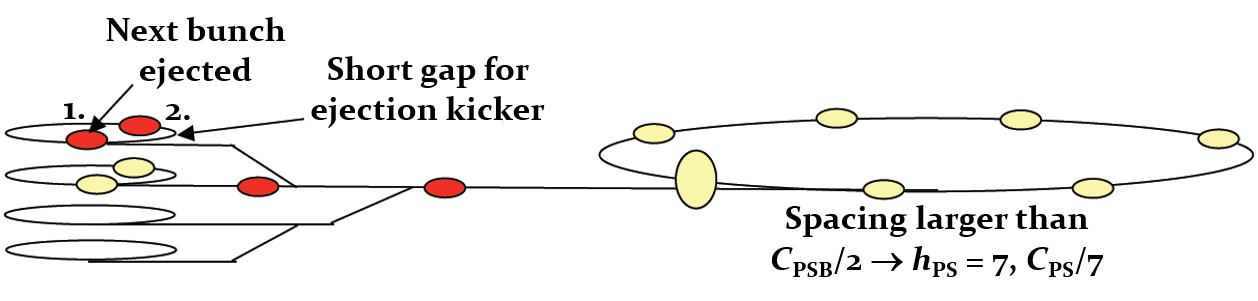}
	\caption{Transfer scheme of pairs of bunches from each PSB ring. In the double-harmonic RF configuration, the bunches are slightly displaced in azimuth to achieve the desired spacing suitable for injection into the PS. Illustration courtesy of C. Carli, CERN.}
	\label{fig:SingleBatchTransferCarliSketch}
\end{figure}

As illustrated in Fig.~\ref{fig:SingleBatchTransferCarliSketch} it is then sufficient to trigger the ejection and injection kickers such that both bunches from each ring are extracted according to the indicated order.
The ejection kicker of the PSB is triggered during the short spacing to obtain two bunches spaced by $327\unsty{ns}$. This special scheme, based on the special, non-linear RF voltage shape, has been used operationally to save time during fillings of the Large Hadron Collider (LHC) during a few years.

\subsection{Long and short bunches simultaneously}

\begin{figure}[htb]
	\centering
	\includegraphics[width=.29\linewidth]{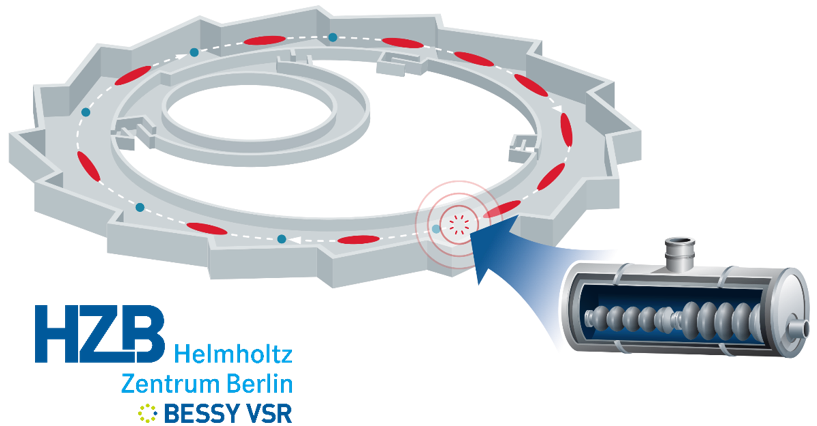}
	\includegraphics[width=.7\linewidth]{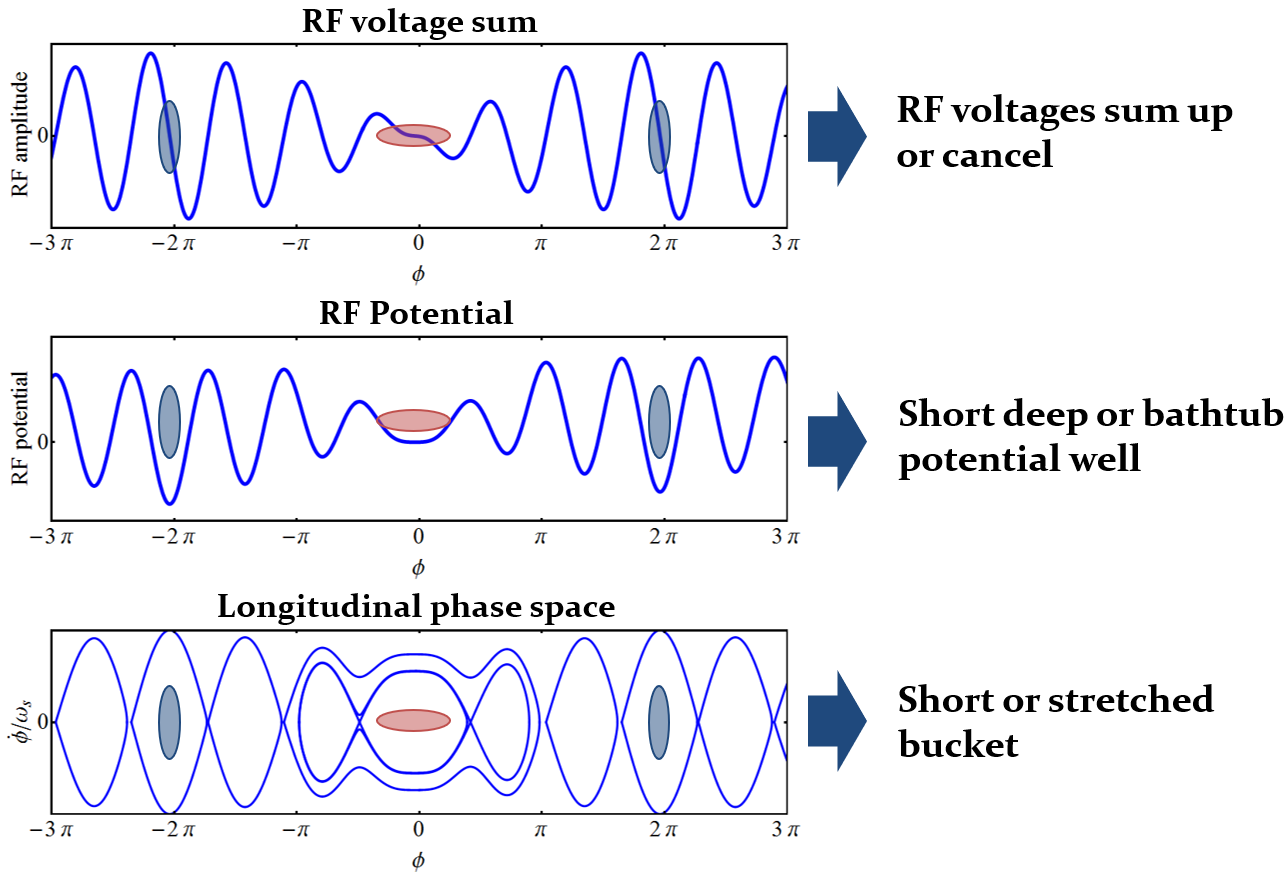}
	\caption{RF voltage (top), potential (middle) and longitudinal phase space (bottom) for the simultaneous long (blue) and and short (red) mode, as proposed for the electron storage ring BESSY. Super-conducting RF cavities are the only technique to provide sufficiently large RF voltages. Illustration (left) courtesy of BESSY.}
	\label{fig:LongShortBunchVoltagePotentialLongitudinalPhaseSpace}
\end{figure}

Conventional electron storage rings can either be operated with long or short bunches. This mode of operation defines the main types of experiments which can be performed with the synchrotron radiation from these rings. An innovative scheme, based on non-linear RF voltages has been proposed for the storage ring BESSY to produce long and short bunches simultaneously~\cite{bib:hzb2015}. Adding a higher harmonic RF voltage at 3 ($1.5\unsty{GHz}$) and 3.5 times ($1.75\unsty{GHz}$) the frequency of the fundamental RF system at $500\unsty{MHz}$ yield an effective RF voltage as illustrated in Fig.~\ref{fig:LongShortBunchVoltagePotentialLongitudinalPhaseSpace} (top).
Depending on the azimuth of the bunch the RF voltage either sum up or compensate each other at the position of the bunch. The resulting RF potential~(Fig.~\ref{fig:LongShortBunchVoltagePotentialLongitudinalPhaseSpace}, middle) has either short and deep potential wells for short bunches, or when the voltage of the different RF harmonics cancels at the bunch position, a rather long bathtub like potential well forms, which stretches the bunches in these buckets.

The shape of the buckets~(Fig.~\ref{fig:LongShortBunchVoltagePotentialLongitudinalPhaseSpace}, bottom) illustrates the longitudinal phase space featuring short buckets with large momentum acceptance and comparably long buckets with small bucket height. This particular mode of operations allows to virtually combine two storage rings with different beam parameters into a single one able to simultaneously serve two classes of experiments thanks to this clever application of non-linear RF voltages.

\section{Longitudinal stability with multi-harmonic RF systems}

As shown in Sec.~\ref{sec:LongitudinalBeamDynamics} each particle in a bunch executes synchrotron oscillations around the synchronous particle at the centre of the bucket. This causes a continues exchange of phase and energy offset. As long as the surrounding trajectory of a bunch matches with the longitudinal distribution, the macroscopic bunch distribution and profile remains unchanged.

The distribution of the synchrotron frequencies depends on the non-linearity of the RF voltage. Only in the special case of a linear acceleration voltage, as illustrated in Fig.~\ref{fig:VoltagePotentialPhaseSpaceLinearVoltage}, the frequency of the synchrotron oscillation is identical for all particles.

\begin{figure}[htb]
	\centering\includegraphics[width=.8\linewidth]{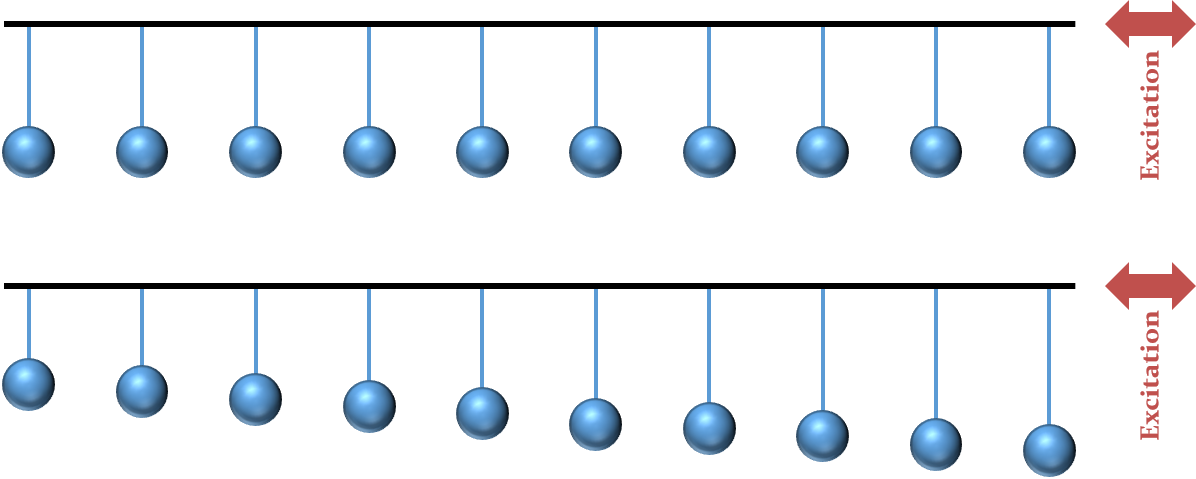}
	\caption{Ensemble of oscillators represented by pendulums attached to a horizontal bar. The pendulums are excited by a horizontal movement of the bar, as indicated by the red arrows. When all pendulums have the same length (top), all of them can be excited coherently at their resonance frequency. Introducing a spread of individual resonance frequencies makes the coherent excitation impossible.}
	\label{fig:PedulumEquivalent}
\end{figure}

A bunch can generally be described as an ensemble of harmonic oscillators, each of which with its own resonant frequency. The entire bunch becomes longitudinally unstable when coherent oscillations are excited, which means driving a large fraction of the individual particles simultaneously at their resonance frequency. This idea is illustrated in Fig.~\ref{fig:PedulumEquivalent}.
A number of pendulums are hanging at a horizontal bar which is excited by a horizontal movement. When all oscillators have the same resonance frequency~(Fig.~\ref{fig:PedulumEquivalent}, top), they can be easily excited coherently by a movement at that specific frequency. The entire system becomes very unstable. This behavior changes significantly when the pendulums all have slightly different resonance frequencies. Whichever excitation frequency is chosen to drive them, only a small number of pendulums will start to oscillate and a coherent motion of the entire cannot develop. Macroscopically the set of oscillators remains perfectly stable ~(Fig.~\ref{fig:PedulumEquivalent}, top). This damping due to a spread of resonance frequencies of individual oscillators is also referred to as Landau damping~\cite{bib:hofmann2006}. The wider the spread, the more stable the ensemble of oscillators will be macroscopically.

The same mechanism applies to the stability of a bunch in a bucket. The larger the spread of the resonance frequencies, the synchrotron frequencies of the individual particles, the more stable the bunch will generally be against coherent excitation. To judge the longitudinal stability of a bunch, the distribution of synchrotron frequencies of the particles in the bunch is hence most relevant.

\subsection{Synchrotron frequency distribution}

The synchrotron frequency is the inverse of the time it takes a particle to perform one complete oscillation on a closed trajectory in the longitudinal phase space. The average phase velocity for a given trajectory in the longitudinal phase space can be calculated introducing the action angle variable, $J(H)$. For a particle on a given trajectory, $\dot{\phi} (\phi)$ defined by the constant value of its Hamiltonian, the corresponding action angle becomes
\begin{equation}
    J(H) = \frac{1}{2 \pi \omega_\mathrm{S}} \oint \dot{\phi} (\phi) \, d\phi \, .
\end{equation}
The angular frequency, the synchrotron frequency of the particle on the given trajectory is then calculated according to
\begin{equation}
    \omega (H) = \frac{d}{dJ} H \, ,
\end{equation}
and the general expression for the synchrotron frequency can be expressed as
\begin{equation}
    \frac{ \omega (H) }{ \omega_\mathrm{S} } = \cfrac{ \sqrt{2} \pi }{ \displaystyle \int_{\phi_\mathrm{l}}^{\phi_\mathrm{u}} \cfrac{1}{\sqrt{H - W(\phi) } } \, d\phi } \,  .
    \label{eqn:GeneralSynchrotronFrequencyDistribution}
\end{equation}
The integration in the denominator is computed from the lower boundary, $\phi_\mathrm{l}$ to the upper boundary, $\phi_\mathrm{u}$ of the trajectory with the value $H$ of the Hamiltonian, $\dot{\phi} (\phi_\mathrm{l})$ = $\dot{\phi} (\phi_\mathrm{u}) = 0$.

\subsubsection{Linear RF voltage}

As a simple crosscheck, the synchrotron frequency distribution can be derived starting from the Hamiltonian Eq.~(\ref{eqn:HamiltonianLinearVoltageNormalized}). The integral in the denominator of Eq.~(\ref{eqn:GeneralSynchrotronFrequencyDistribution}) becomes
\begin{equation*}
    \int_{\phi_\mathrm{l}}^{\phi_\mathrm{u}} \cfrac{1}{\sqrt{H - W(\phi) } } \, d\phi = \sqrt{2} \int_{\phi_\mathrm{l}}^{\phi_\mathrm{u}} \frac{1}{\sqrt{\phi^2_\mathrm{u} - \phi^2}} \, d\phi = 2 \sqrt{2} \left[ \arctan \frac{\phi}{\sqrt{\phi^2_\mathrm{u} - \phi^2}} \right]_0^{\phi_\mathrm{u}} = \sqrt{2} \pi \, .
\end{equation*}
The trajectories in the linear bucket are of course symmetric with respect to $\phi = 0$ and $\Delta E$, as illustrated in Fig.~\ref{fig:VoltagePotentialPhaseSpaceLinearVoltage} and hence $H_\mathrm{trajectory} = H(\phi = \phi_\mathrm{l}, \dot{\phi}/\omega_\mathrm{S}^2 = 0) = H(\phi = \phi_\mathrm{u}, \dot{\phi}/\omega_\mathrm{S}^2 = 0) = 1/2 \cdot \phi_\mathrm{u}^2$ is applied to obtain the integral. As expected from the linear Eqs.~(\ref{eqn:firstHamiltonEquationLinearVoltage}) and (\ref{eqn:secondHamiltonEquationLinearVoltage}), all particles have the same synchrotron frequency $\omega(H) = \omega_\mathrm{S}$ when the RF voltage is linear. This also applies to the synchrotron frequency in the centre of the conventional bucket generated by a sinusoidal RF voltage.

\subsubsection{Sinusoidal RF voltage}

\begin{figure}[htb]
	\centering\includegraphics[width=0.8\linewidth]{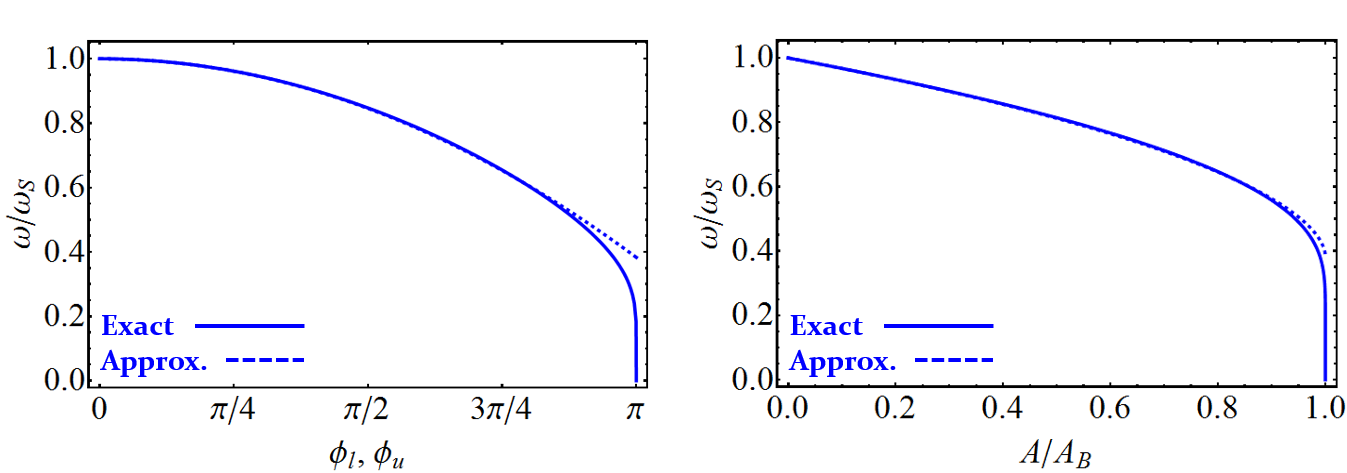}
	\caption{Synchrotron frequency distribution in a single-harmonic stationary bucket versus maximum phase excursion of the trajectory (left) and versus the normalized area, $A/A_\mathrm{B}$ encircled by the trajectory. The total bucket area is $A_\mathrm{B}$. The dashed line shows the approximation according to Eq.~(\ref{eqn:synchrotronFrequencyDistributionStationaryBucket}).}
	\label{fig:StationaryBucketSynchrotronFrequencyDistribution}
\end{figure}

For the more general case of a sinusoidal RF voltage but no acceleration, $\phi_\mathrm{S} = 0$, the exact potential is proportional to $1-\cos \phi = 2 \sin^2 (\phi/2)$. Again, profiting from the symmetry of the trajectories around the bucket centre for the stationary bucket, the denominator integral of Eq.~(\ref{eqn:GeneralSynchrotronFrequencyDistribution}) can be expressed as
\begin{equation}
    \int_{\phi_\mathrm{l}}^{\phi_\mathrm{u}} \cfrac{1}{\sqrt{H - W(\phi) } } \, d\phi = \sqrt{2} \int_0^{\phi_\mathrm{u}} \frac{1}{\sqrt{ \sin^2 (\phi_\mathrm{u}/2) - \sin^2 (\phi/2) }} \, d\phi \, ,
\end{equation}
where $\phi$, $\phi_\mathrm{u}$ denote phase offsets with respect to the bucket center. The Hamiltonian has been replaced by the maximum phase excursion $\phi_\mathrm{u}$ of the trajectory under consideration. With the substitution 
\begin{equation}
    \xi(\phi) = \arcsin \left[ \frac{ \sin (\phi/2) }{ \sin (\phi_\mathrm{u}/2)} \right] 
\end{equation}
this integral reduces to the conventional form of the elliptic integral of the first kind 
\begin{equation*}
    K[\sin (\phi_\mathrm{u}/2)] = \int_0^{\pi/2} \frac{1}{\sqrt{1- \sin^2(\phi_\mathrm{u}/2) \sin^2 \xi}} \, d \xi \, .
\end{equation*}
The synchrotron frequency of a particle on a trajectory with a maximum phase excursion of $\pm \phi_\mathrm{u}$ is expressed in the form of
\begin{equation}
    \frac{ \omega(\phi_\mathrm{u}) }{ \omega_\mathrm{S} } = \frac{ \pi }{ 2 K [ \sin (\phi_\mathrm{u}/2)] } \simeq 1 - \frac{\phi_\mathrm{u}^2}{16} \, .
    \label{eqn:synchrotronFrequencyDistributionStationaryBucket}
\end{equation}
While the synchrotron frequency remains rather constant at the centre of the bucket (Fig.~\ref{fig:StationaryBucketSynchrotronFrequencyDistribution}),
it rapidly decreases towards larger synchrotron oscillation amplitudes. At the separatrix the period of the synchrotron oscillations diverges to infinity. A particle on that trajectory would move very slowly towards the unstable fixed point in-between buckets, but never reach it.

As an extension of Eq.~(\ref{eqn:synchrotronFrequencyDistributionStationaryBucket}) an analytic approximation for the synchrotron frequency distribution versus $\Delta \phi_\mathrm{u} = \phi_\mathrm{u} - \phi_\mathrm{S}$ in an accelerating bucket with the stable phase $\phi_\mathrm{S}$ can be obtained as~\cite{bib:zotter1981, bib:zotter2012}
\begin{equation}
    \frac{ \omega(\Delta \phi_\mathrm{u}) }{ \omega_\mathrm{S} } \simeq 1 - \frac{1 + 2/3 \sin^2 \phi_\mathrm{S}}{16 \, (1-\sin^2 \phi_\mathrm{S})} \Delta \phi_\mathrm{u}^2\, .
\end{equation}
For more complicated RF voltages, like double-harmonic RF systems, analytical approximates can only be calculated for few limited cases~\cite{bib:bramham1977}. However, semi-analytical evaluation Eq.~(\ref{eqn:GeneralSynchrotronFrequencyDistribution}) is most useful.

\subsection{Longitudinal stability and synchrotron frequency spread}

\begin{figure}[htb]
	\centering\includegraphics[width=0.8\linewidth]{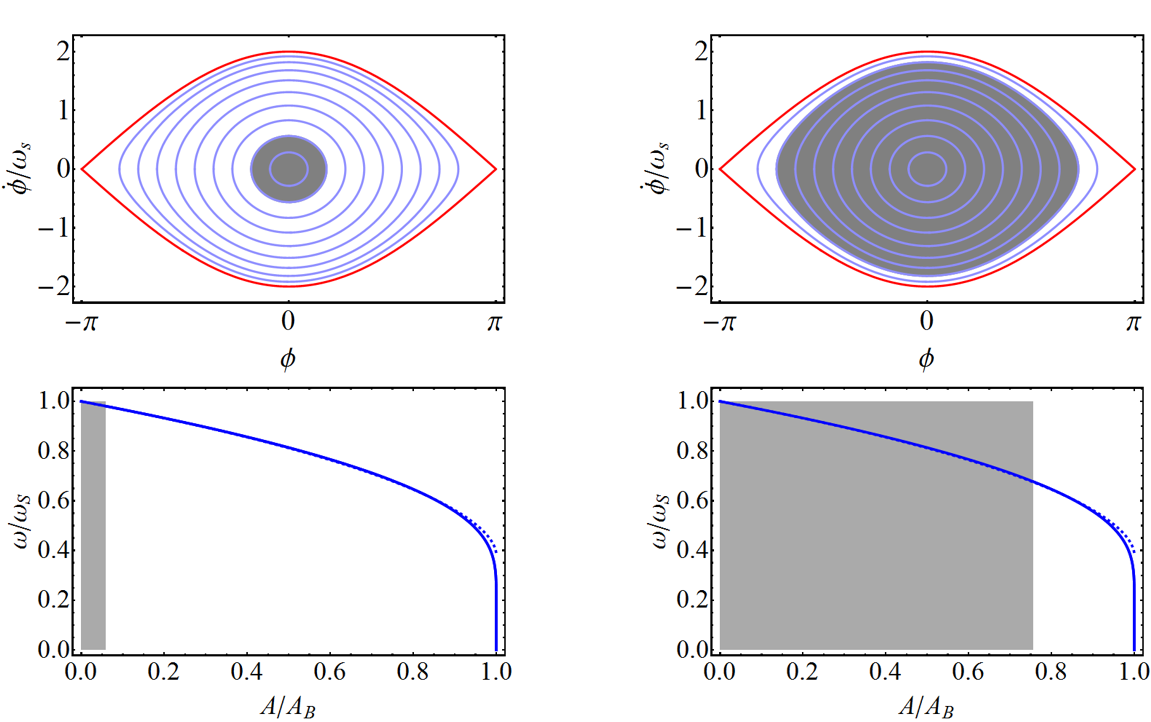}
	\caption{Comparison of longitudinal phase space and synchrotron frequency spread for different bucket filling factors. When the bucket area is large compared to the longitudinal emittance (gray shaded area), the synchrotron frequency spread is small (left). Increasing the bucket filling factor also increases the synchrotron frequency spread and improved the longitudinal stability.}
	\label{fig:StationaryBucketSynchrotronFrequencyDistributionBunch}
\end{figure}

As illustrated in Fig.~\ref{fig:PedulumEquivalent}, the larger the spread of the resonance frequencies of an ensemble of oscillators, the more stable the entire system will be against any single-harmonic excitation. For a bunch in a stationary bucket this is sketched in Fig.~\ref{fig:StationaryBucketSynchrotronFrequencyDistributionBunch}.
The area, $A$, encircled by a trajectory in the longitudinal phase space is equivalent to the total longitudinal emittance, $\varepsilon_\mathrm{l}$ of a matched bunch. Choosing a large RF voltage and a large bucket area for a small emittance bunch~(Fig.~\ref{fig:PedulumEquivalent}, left) may suggest important margins against the loss of particles from the bucket. The gray shaded area indicates the fraction of the bucket occupied by particles. In reality, due to the small synchrotron frequency spread, such a small bunch is prone to longitudinal instabilities. When decreasing RF voltage and bucket area or increasing the longitudinal emittance the spread of synchrotron frequencies covered by the bunch becomes wider (right). As a consequence, such a bunch is more resistant against any excitation. Simply reducing the RF voltage during the acceleration cycle, to avoid that the synchrotron frequency spread becomes too small, is hence an efficient means to improve longitudinal stability in hadron synchrotrons. This technique is also referred to as constant bucket area acceleration.

\subsection{Stability in double-harmonic RF systems}

To improve the longitudinal stability beyond the range achievable by voltage reduction, the non-linearity of the RF voltage can be increased by adding a second RF system. In most cases it operates at an integer multiple of the fundamental RF system, and the normalized RF voltage can be written as
\begin{equation}
    g(\phi) = \frac{1}{V} \left[ \, V_1 \sin \phi + V_2 \sin (h_2/h_1 \phi + \phi_1) \, \right] \, ,
\end{equation}
where $V_{1,2}$ and $h_{1,2}$ are again the voltage amplitudes and harmonic numbers of the two RF systems. The relative phase between both of them is $\phi_1$. Figure~\ref{fig:doubleHarmonicBunchShorteningLengtheningModeExample} introduces the two main modes of operation. When both RF voltages are in phase at the centre of the bunch (bunch shortening mode), the effective gradient increases and the bunch length decreases. In the opposite case (bunch lengthening) the voltage gradient is reduced or even inverted and hence the bunch is stretched.

\subsubsection{Bunch shortening mode}
\label{sec:bunchShorteningMode}

\begin{figure}[htb]
	\centering
	\includegraphics[width=0.35\linewidth]{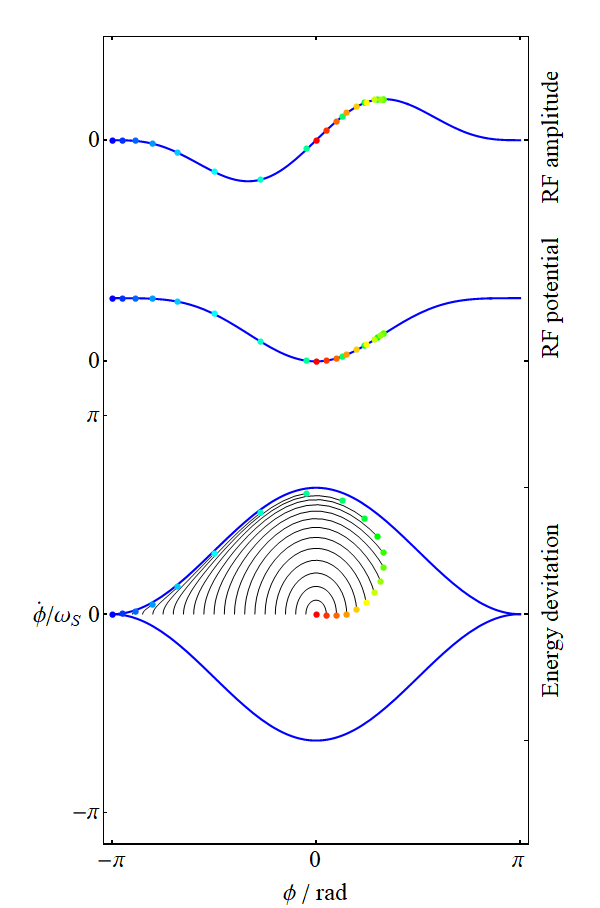}
	\hspace*{3em}
	\includegraphics[width=0.35\linewidth]{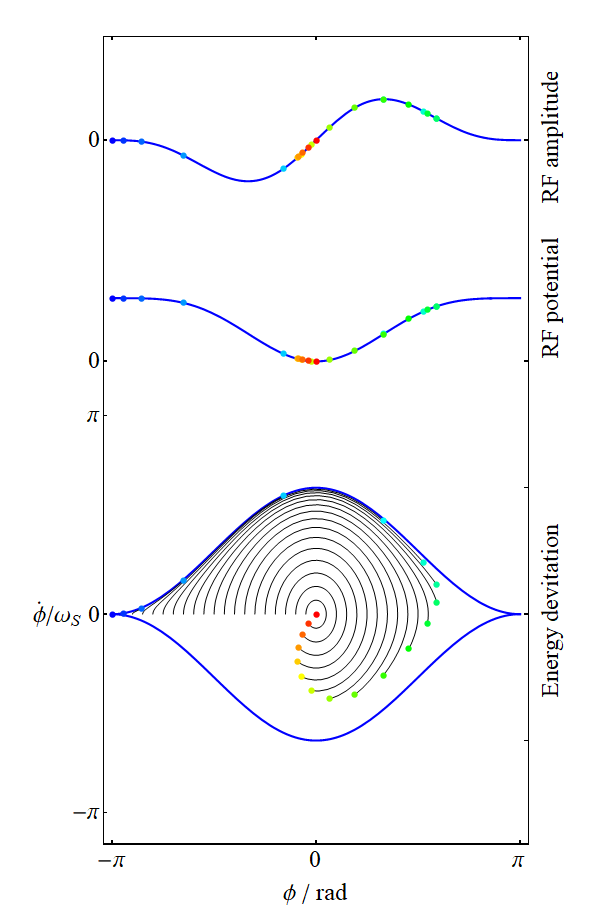}
	\caption{Amplitude (top), resulting potential (middle) and normalized longitudinal phase space (bottom) in coordinates of $\phi$ and $\dot{\phi}/\omega_\mathrm{S}$ for the case of two RF systems in phase with a harmonic and voltage ratio of $V_2/V_1 = 1/2$  and $h_2/h_1 = 2.$}
	\label{fig:DoubleHarmonicBunchShorteningPhaseSpaceConstructionAllRatio2}
\end{figure}

\begin{figure}[htb]
	\centering\includegraphics[width=0.4\linewidth]{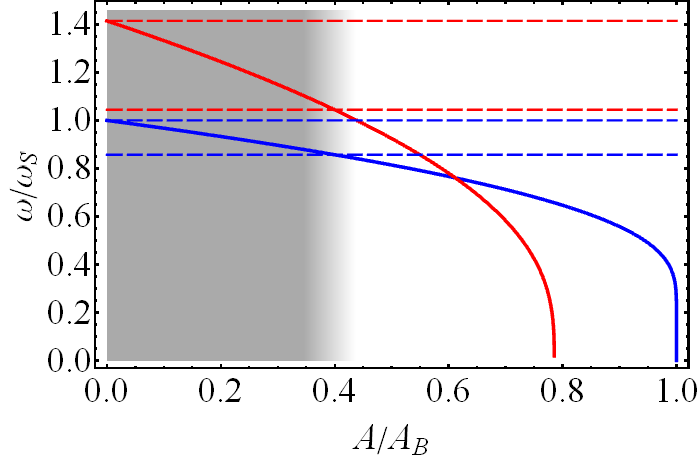}
	\caption{Synchrotron frequency distribution versus the bucket filling factor for a single- (blue) and double-harmonic RF voltage (red, $V_2/V_1 = 1/2$  and $h_2/h_1 = 2$).}
	\label{fig:DoubleHarmonicBunchShorteningPhaseSpaceConstructionAllSynchrotronFrequencySpread}
\end{figure}

The RF amplitude, potential and resulting longitudinal phase space are shown in Fig.~\ref{fig:DoubleHarmonicBunchShorteningPhaseSpaceConstructionAllRatio2} for a set of test particles launched with phase but no energy error. Both RF systems are in phase at the centre of the stationary bucket.
After half a period of the synchrotron frequency for the particles in the centre of the bucket (left), the particles towards the separatrix massively lag behind, indicating a large spread of synchrotron frequencies. This effect is even stronger after a full synchrotron frequency period (right). A comparison of the synchrotron frequency distribution between single- and double-harmonic RF voltage versus the bucket filling factor is plotted in Fig.~\ref{fig:DoubleHarmonicBunchShorteningPhaseSpaceConstructionAllSynchrotronFrequencySpread}.
It confirms that the synchrotron frequency spread for a typical bucket filling ratio, $\varepsilon_\mathrm{l}/A_\mathrm{B}$ of about $40\unsty{\%}$ is more than doubled.

\begin{figure}[htb]
	\centering
	\includegraphics[width=0.35\linewidth]{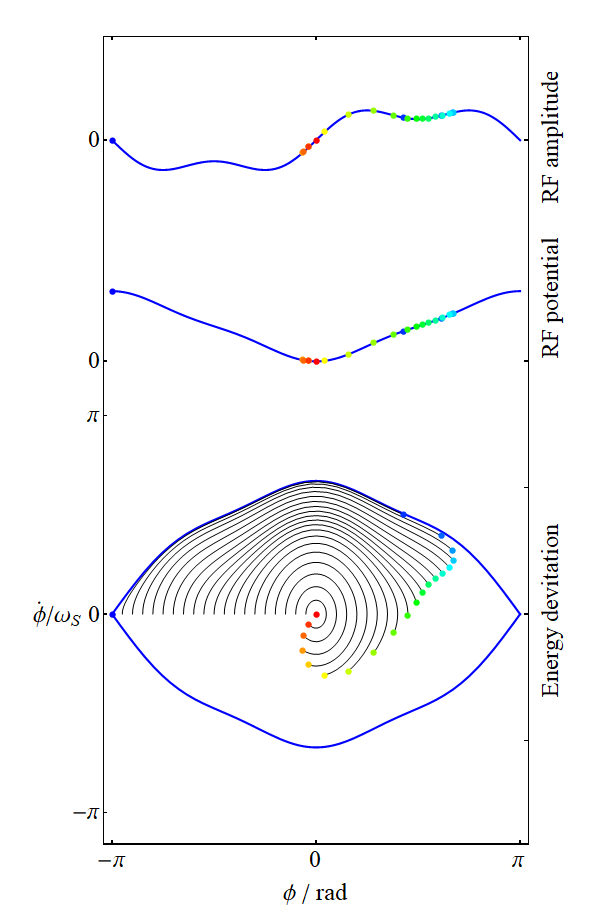}
	\hspace*{3em}
	\includegraphics[width=0.35\linewidth]{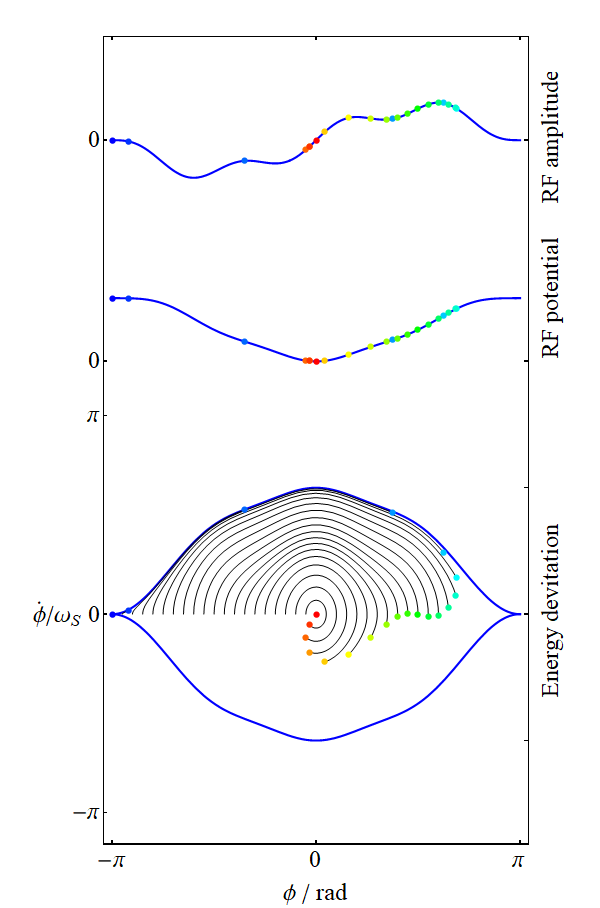}
	\caption{Amplitude (top), resulting potential (middle) and normalized longitudinal phase space (bottom) in coordinates of $\phi$ and $\dot{\phi}/\omega_\mathrm{S}$ for the case of two RF systems in phase with a harmonic and voltage ratios of $V_2/V_1 = 1/3$ and $h_2/h_1 = 3$ (left), $V_2/V_1 = 1/4$ and $h_2/h_1 = 4$ (right). The plots show the situation after one period of the synchrotron frequency in the center of the bucket.}
	\label{fig:DoubleHarmonicPhaseSpaceConstructionComparisons}
\end{figure}

\begin{figure}[htb]
	\centering
	\includegraphics[width=0.329\linewidth]{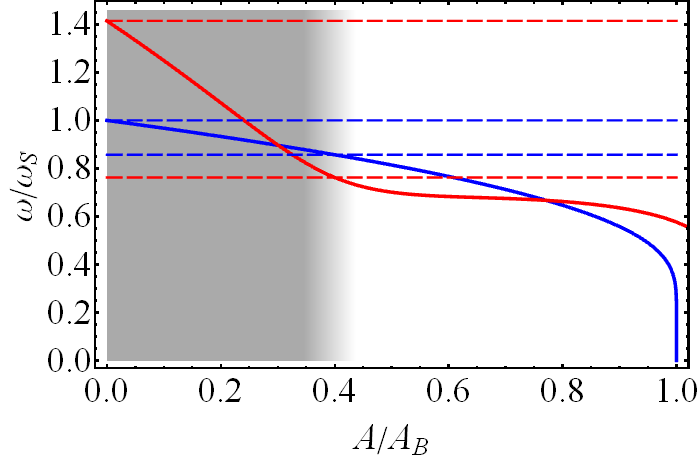}
	\hspace*{4em}
	\includegraphics[width=0.329\linewidth]{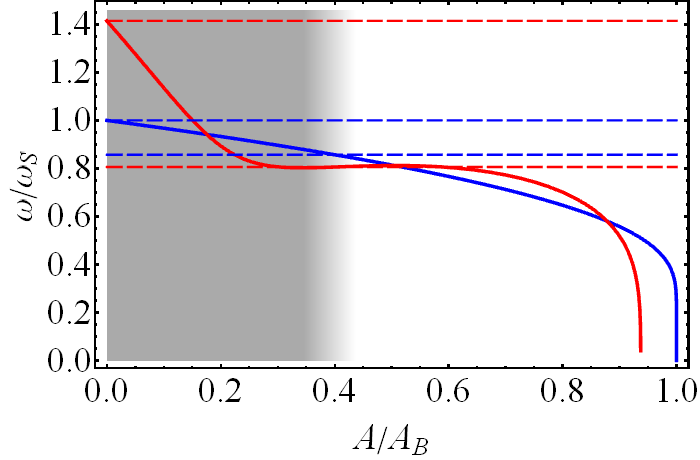}
	\caption{Synchrotron frequency distribution versus the bucket filling factor for the two cases shown in Fig.~\ref{fig:DoubleHarmonicPhaseSpaceConstructionComparisons}. The harmonic and voltage ratios in these examples are $V_2/V_1 = 1/3$ and $h_2/h_1 = 3$ (left), $V_2/V_1 = 1/4$ and $h_2/h_1 = 4$ (right).}
	\label{fig:DoubleHarmonicPhaseSpaceConstructionComparisonsSynchrotronFrequencySpread}
\end{figure}

This effect can even be increased by choosing a higher harmonic ratio, as illustrated in Fig.~\ref{fig:DoubleHarmonicPhaseSpaceConstructionComparisons}
for the harmonic number ratios of $1/3$ and $1/4$ in the centre of the bucket. Clearly the lag of particles with phase of momentum offset with respect to the synchronous particle increases quickly. Approaching however, the separatrix the motion in longitudinal phase space get faster again. The effect also manifests in the synchrotron frequency distribution~(Fig.~\ref{fig:DoubleHarmonicPhaseSpaceConstructionComparisonsSynchrotronFrequencySpread}, red),
again compared to the single-harmonic case (blue). A higher-harmonic RF system a $h_2/h_1 = 4$ is operated very successfully in Super Proton Synchrotron (SPS) at CERN to stabilize the high-intensity beam for the LHC~\cite{bib:shaposhnikova1998,bib:shaposhnikova2005}. The thresholds of longitudinal instability are increased by a large factor.

Although the spread in the bunch centre grows further with the harmonic number ratio, the synchrotron frequency function, $\omega/\omega_\mathrm{S}$ may not be monotonic develop saddle points or even invert its slope towards the outer part of the bucket. In such local regions with no gradient or inversion, a large number of particles will again have the same synchrotron frequency, making it very susceptible to any excitation. The longitudinal stability in this case becomes even worse than the single-harmonic case which, even in the bunch centre, guarantees a small frequency spread. Populating these regions of the bucket with particles must be avoided by either keeping the longitudinal emittance small enough or reducing the voltage of the second harmonic RF system.

To longitudinally stabilize a beam with a double-harmonic RF voltage, the harmonic number and voltage ratios must therefore be carefully chosen. The larger the harmonic ratio, the bigger the synchrotron frequency spread for the same voltage, but at the expense of potentially reducing the usable fraction of the bucket area.

\subsubsection{Bunch lengthening mode}

The configuration where the two RF system are in counter-phase at the bunch position is very beneficial to reduce the peak current, and hence the longitudinal space charge effect in the stretched bunch as shown in the example of Fig~\ref{fig:doubleHarmonicBunchLengtheningModeExamplePSB}.

\begin{figure}[htb]
	\centering
	\includegraphics[width=0.35\linewidth]{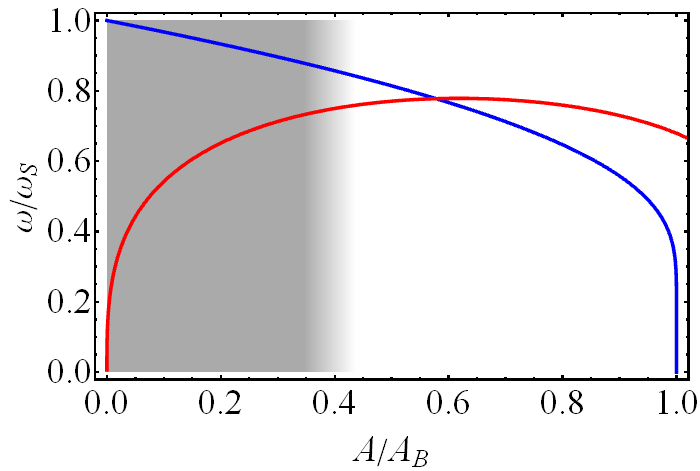}
	\hspace*{4em}
	\includegraphics[width=0.35\linewidth]{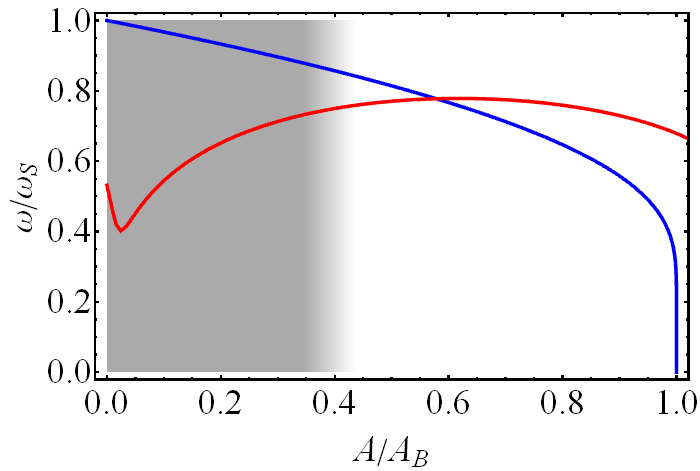}
	\caption{Synchrotron frequency distribution versus the bucket filling factor in a double-harmonic RF system with both voltages in counter phase ($V_2/V_1 = 1/2$ and $h_2/h_1 = 2$. In case of a perfect relative phase (left), the synchrotron frequency vanishes at the bunch centre. A small, unavoidable relative phase error of only $\Delta \phi = 5^0$ is however sufficient to change the synchrotron frequency distribution significantly (right).}
	\label{fig:DoubleHarmonicBunchLengtheningPhaseSpaceConstructionAllSynchrotronFrequencySpread}
\end{figure}

At first sight also the synchrotron frequency distribution as plotted in Fig.~\ref{fig:DoubleHarmonicBunchLengtheningPhaseSpaceConstructionAllSynchrotronFrequencySpread} (left) seems very attractive to stabilize a bunch.
Close to the bunch centre particles move very slowly with respect to the synchronous particle. The synchrotron frequency then grows for larger bucket filling ratios and decreases again when approaching the separatrix. Although the spread is extremely large in this ideal case, a small phase error between both RF systems has a detrimental impact. Already with a small residual phase error~(Fig.~\ref{fig:DoubleHarmonicBunchLengtheningPhaseSpaceConstructionAllSynchrotronFrequencySpread}, right), unavoidable in a real accelerator, the slope flips and regions with the same synchrotron frequency are found in different part of the central part of the bunch. Such a synchrotron frequency distribution is not beneficial for longitudinal stability. In bunch shortening the synchrotron frequency distribution is insensitive to such small phase errors due to both voltages adding up in phase at the bucket centre.

\begin{figure}[htb]
	\centering\includegraphics[width=\linewidth]{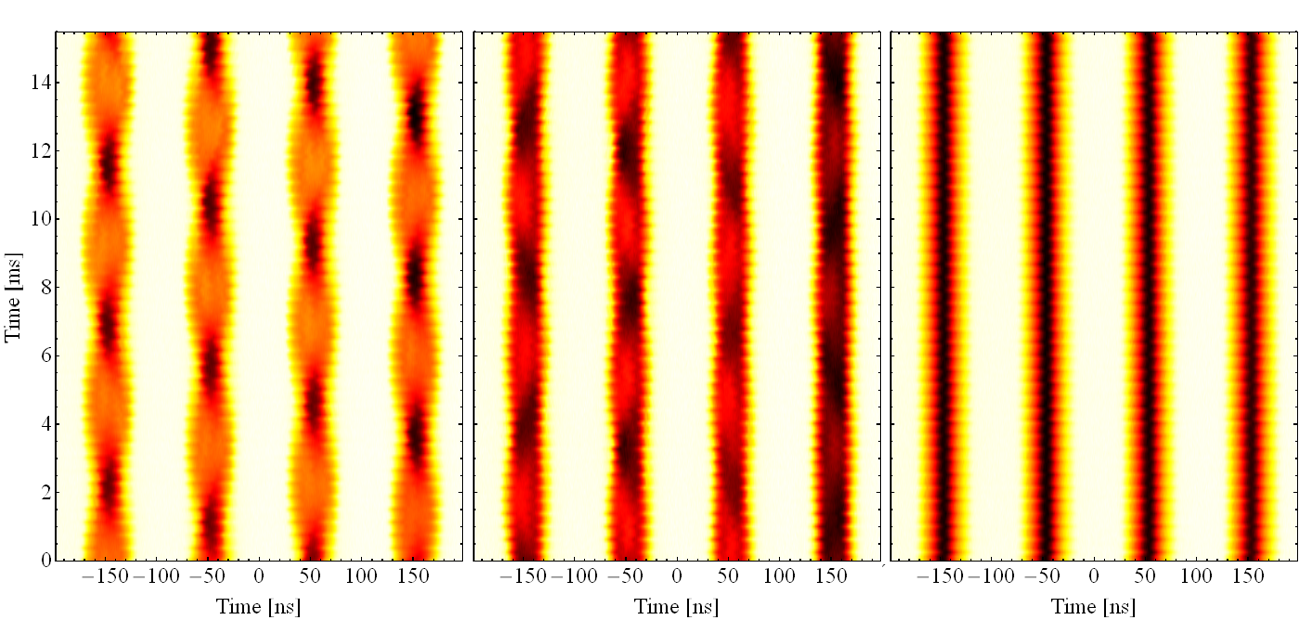}
	\caption{Measured evolution of the bunch profile in the PS during $15\unsty{ms}$ at the flat-top in the PS at CERN. Only four bunches of a train of $18$ are shown. In the single-harmonic case (left) strong quadrupole oscillations are observed, which partially persist when adding voltage at the fourth harmonic in counter phase to stretch the bunches (middle). With the voltage at the fourth harmonic in phase, the oscillations are entirely damped (right).}
	\label{fig:SingleDoubleHarmonicStabilityPSExample}
\end{figure}

The longitudinal stability at the flat-top in the CERN PS has been checked with beam for the three different case of single-harmonic RF at $h_1=21$, adding RF voltage at $h_2=84$ ($h_2/h_1 = 4$) in counter phase and in phase.
As can be seen in the mountain range plot of four exemplary bunches in Fig.~\ref{fig:SingleDoubleHarmonicStabilityPSExample},
the bunches exhibit strong bunch length oscillations (left). A constant bunch-to-bunch phase advance is observed, indicating a coupled-bunch oscillation. With the higher-harmonic RF system in counter-phase longitudinal stability is improved marginally. When flipping the phase of the higher harmonic RF voltage, the oscillations entirely disappear and the bunches are stable.

Although the longitudinal peak density is highest when both RF systems are in phase, the bunches are most stable, with no sign of quadruplolar coupled-bunch oscillations. As has been shown above, this stability is achieved thanks to the enlarged synchrotron frequency spread introduced by an intentional non-linearity caused by the second RF system.

\section{Conclusion}

Multi-harmonic RF systems allow to manipulate the longitudinal beam parameter like bunch length and spacing. The reduction of space charge effects by reducing the peak line density in proton synchrotrons or the simultaneous operation of short and long bunches in an electron storage ring are examples of the wide range of applications. 

The non-linearity of the RF voltage applied to a charged particle beam in accelerators is essential to keep high-intensity beams longitudinally stable. In many cases additional non-linearity in the form of multi-harmonic RF systems is intentionally implemented, to manipulate the longitudinal beam parameters like bunch length or distance between bunches. Of equal importance is the improvement of the longitudinal beam stability which is achieved with multi-harmonic RF systems.

The most common case, double-harmonic RF, is installed in most high-intensity electron and proton synchrotrons. With appropriately chosen parameters, the non-linearity due to a high-harmonic RF system induces a wider spread of synchrotron frequencies of the particles in a bunch. The associated gain in longitudinal beam stability is significant and allows to extend the performance reach in terms of bunch intensity and longitudinal beam density well beyond the possibilities with a conventional, single-harmonic RF system.

\section{Acknowledgement}

The author would like to thank Wolfgang H\"ofle, Andreas Jankowiak, Erk Jensen, Danilo Quartullo, Markus Ries, Elena Shaposhnikova and Frank Tecker for discussions and material provided in preparation of this course. Thomas Bohl is acknowledged for his valuable corrections of the manuscript and numerous suggestions for improvements.

\end{document}